\documentclass[conference,10pt]{IEEEtran}
\usepackage{epsfig,rotating,setspace,latexsym,amsmath,epsf,amssymb,amsfonts,bm,theorem, epstopdf,cite,authblk, bbm, caption, subcaption, multirow, enumitem}
\usepackage[ruled,vlined]{algorithm2e}

\usepackage{color}

\IEEEoverridecommandlockouts
\allowdisplaybreaks

\begin{document}
\title{Adversarial Attacks on LoRa Device Identification and Rogue Signal Detection with Deep Learning}	
	
\author[]{Yalin E. Sagduyu and Tugba Erpek}

\affil[]{\normalsize  Virginia Tech, Arlington, VA, USA \\ Email: \{ysagduyu, terpek\}@vt.edu}
	
\maketitle
\begin{abstract}
Low-Power Wide-Area Network (LPWAN) technologies, such as LoRa, have gained significant attention for their ability to enable long-range, low-power communication for Internet of Things (IoT) applications. However, the security of LoRa networks remains a major concern, particularly in scenarios where device identification and classification of legitimate and spoofed signals are crucial. This paper studies a deep learning framework to address these challenges, considering LoRa device identification and legitimate vs. rogue LoRa device classification tasks. A deep neural network (DNN), either a convolutional neural network (CNN) or feedforward neural network (FNN), is trained for each task by utilizing real experimental I/Q data for LoRa signals, while rogue signals are generated by using kernel density estimation (KDE) of received signals by rogue devices. Fast Gradient Sign Method (FGSM)-based adversarial attacks are considered for LoRa signal classification tasks using deep learning models. The impact of these attacks is assessed on the performance of two tasks, namely device identification and legitimate vs. rogue device classification, by utilizing separate or common perturbations against these signal classification tasks. Results presented in this paper quantify the level of transferability of adversarial attacks on different LoRa signal classification tasks as a major vulnerability and highlight the need to make IoT applications robust to adversarial attacks.
\end{abstract}

\section{Introduction}
\emph{LoRa communication} technology offers long-range connectivity, low power consumption, extended battery life, and scalability, making it a cost-effective solution for diverse Internet of Things (IoT) applications based on Low Power Wide Area Networks (LPWAN) such as asset tracking, industrial automation, surveillance, smart city, smart home, and supply chain management \cite{bor2016lora, sinha2017survey, zourmand2019internet, LoRatactical1}. While LoRa offers several advantages, it also poses various \emph{security challenges}. LoRa networks may be vulnerable to unauthorized access such that adversaries can potentially gain access to the network, intercept or manipulate data, and disrupt the communication between devices. LoRa signals can be intercepted and eavesdropped over the air. In replay attacks, adversaries may capture and retransmit legitimate LoRa signals to deceive the network, or they may spoof or impersonate legitimate devices of a LoRa network to gain unauthorized access, inject malicious data, or disrupt network operations. To that end, it is essential to characterize the attacks surface for LoRa. 

\emph{Machine learning} plays a pivotal role in  wireless signal classification applications within LoRa networks. The unique advantage of \emph{deep learning} lies in its ability to automatically extract high-level features from raw signal data, enabling accurate and efficient classification \cite{erpek2020deep, west2017deep, shi2019deep}. In the context of LoRa, deep neural network (DNN) models such as convolutional neural networks (CNNs) and feed-forward neural networks (FNNs) can effectively differentiate between different types of signals by learning intricate patterns and representations due to the complexity and variability of wireless signals. By training on real-world in-phase and quadrature (I/Q) data for LoRa signals, the DNNs fostered by recent computational advances can capture nuanced signal characteristics, leading to improved classification accuracy \cite{Lora1, robyns2017physical, elmaghbub2021lora, al2021deeplora, shen2022towards, 9621015,9093371}. 

Wireless signals can be analyzed for multiple tasks at a receiver \cite{jagannath2021multi}. In this paper, we consider \emph{two signal classification tasks} for LoRa networks, 
namely the task of distinguishing between devices with RF fingerprinting on the received signals and the task of detecting spoofed signals that mimic the LoRa signal characteristics. \emph{Synthetic signal} generation for spoofed communications by adversaries serves the purpose of deceiving spectrum monitors and potentially bypassing authentication mechanisms \cite{shi2019generative, shi2020generative}. We consider generating these synthetic signals with the \emph{Kernel Density Estimation} (KDE) method. KDE  estimates the probability density function (PDF) of received signals from rogue devices. By modeling the distribution of legitimate LoRa signals, KDE generates synthetic signals that closely mimic the statistical properties of the original signals. This way, adversaries can potentially infiltrate the authentication process and compromise the overall security of the LoRa network.

The complex decision space of wireless signal classification is sensitive to variations in test input samples and therefore vulnerable to \emph{adversarial attacks} (evasion attacks) \cite{sadeghi2018adversarial, kim2021channel, adesina2022adversarial, sagduyu2020wireless}. Adversarial attacks craft small, malicious perturbations in the input signals in test (inference) time to deceive the DNN models, resulting in incorrect device identification or the misclassification of legitimate and rogue signals, and potentially leading to the infiltration of user authentication mechanisms based on RF fingerprinting. 
Addressing the challenges posed by adversarial attacks is crucial to ensure the integrity and security of LoRa networks, enabling trustworthy and dependable wireless communications in various IoT applications.

In this paper, we study \emph{adversarial attacks on wireless signal classification tasks for LoRa}. We consider \emph{untargeted attacks} that aim to disrupt the overall model performance without specifying the target class for each task. We use Fast Gradient Sign Method (FGSM) to generate adversarial inputs by calculating the gradient of the loss function with respect to the input data and perturbing the input in the gradient sign's direction to maximize the model's loss \cite{goodfellow2015explaining}.

\begin{figure}[t]
	\centering
	\includegraphics[width=\columnwidth]{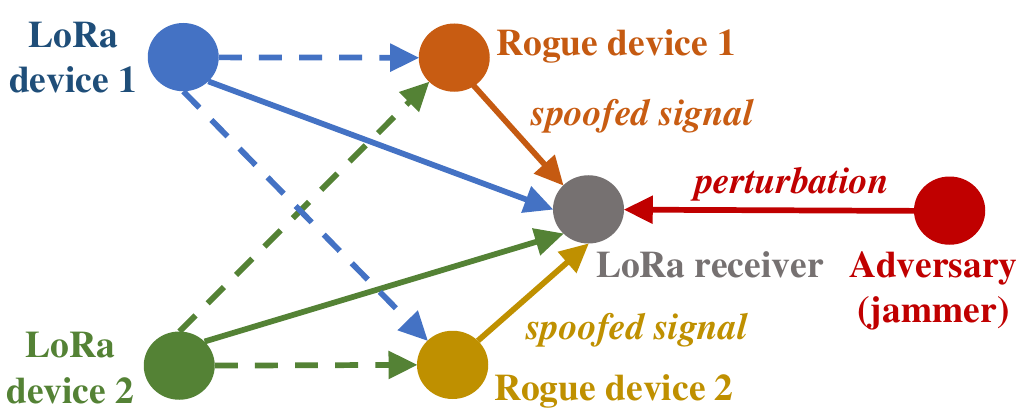}
	\caption{System model.}
	\label{fig:LoRa_system}
\end{figure}

Fig.~\ref{fig:LoRa_system} shows the system model. There are two LoRa devices transmitting and potentially two rogue devices mimicking these transmissions. The LoRa receiver performs two tasks, namely, distinguishing between legitimate transmission from two LoRa devices and distinguishing between legitimate and rogue LoRa device transmissions. There is also an adversary that transmits perturbation signals as part of the adversarial attack. We assess the impact of the adversarial attack on the performance of these two tasks. To that end, we analyze \emph{transferability} of adversarial attacks in the sense that adversarial examples crafted to deceive one task's model can also mislead other task models, even if those models were trained on different datasets. In other words, the adversarial perturbations generated for one model tend to generalize and remain effective across multiple models as the underlying vulnerabilities that are exploited by adversarial attacks are not unique to a specific model but can be present in multiple models due to similar characteristics in their decision boundaries or loss landscapes. The effects of wireless channels and surrogate models on transferability have been studied in \cite{surrogate} for a single task. 

In this paper, we analyze the extent that the adversarial attack for one task transfers to the other task that operates on the same LoRa signal data. We show that the attack performance drops significantly when there is a mismatch between the model under attack and the model for which the perturbation was derived. As a remedy, we utilize a hybrid approach that generates common perturbation by utilizing the gradients of the loss function for multiple classifiers. We show that this attack is highly effective regardless of the classifier model and its type of DNN under attack. 

The rest of the paper is organized as follows. Section~\ref{sec:device} describes LoRA device identification. Section~\ref{sec:rogue} studies detection of rogue LoRa signals. Section~\ref{sec:attack} presents adversarial attacks on LoRa signal classification tasks.
 Section~\ref{sec:conclusion} concludes the paper.

\section{Device Identification from LoRa Signals} \label{sec:device}
We use the I/Q data reported in \cite{elmaghbub2021lora}. This data is collected from real LoRa devices, namely, Pycom IoT devices, as transmitters and software-defined radios (SDRs), namely, USRP B210 receivers, as receivers. The LoRa devices operate in outdoor environment at a center frequency of 915MHz, used for recording the received signals sampled at 1MS/s. The LoRa configuration includes the Raw-LoRa mode, the channel bandwidth of 125kHz, the spreading factor of 7, the preamble of 8, the TX power of 20dBm, and the coding rate of 4/5. The outdoor data in this dataset is collected over 5 consecutive days with 10 transmissions, each of length 20 sec. for each IoT transmitter. The distance between the transmitter and receiver is set as 5m for each experiment.

One task we consider with the use of deep learning is to distinguish between transmissions from two LoRa devices.  
 Each data sample as input to the DNN is of size (2,32) corresponding to 32 I/Q (wireless signal) samples. 5000 samples are generated. 80\% of the samples are used for training and 20\% of them are used for testing. We consider both CNN and FNN classifiers for this task. Their architectures are given in Table~\ref{tab:NNArch}. FNN has 6,522 parameters and CNN has 70,074 parameters. In training of each DNN, we use categorical cross-entropy as the loss to be minimized and Adam \cite{adam} as the optimizer. Table \ref{tab:actual} shows the accuracy of device identification (classification of Device 1 vs. Device 2) using legitimate LoRa device transmissions. In all cases, the accuracy is high. We note that CNN improves performance compared to FNN in terms of average accuracy as well as accuracy of detecting Device 1 or Device 2. 

\begin{table}[ht]
\small
    \centering
    \caption{DNN architectures.}
    \label{tab:NNArch}
    \begin{subtable}[t]{0.5\textwidth}
    \centering
    \caption{DNN type: CNN.}
    \begin{tabular}{l||l}
    \hline
    Layers & Properties \\ \hline
       Conv2D & filters = 32, kernel size = (1,3), \\ & activation function = ReLU\\
       Flatten & -- \\
       Dense & size = 32, activation function = ReLU\\
       Dropout & dropout rate = $0.1$ \\
       Dense & size = 8, activation function = ReLU \\
       Dropout & dropout rate = $0.1$\\
       Dense & size = 2, activation function = SoftMax \\ \hline
    \end{tabular}
    \vspace{0.5cm}
    \end{subtable}
    \begin{subtable}[t]{0.5\textwidth}
    \centering
    \caption{DNN type: FNN.}
    \begin{tabular}{l||l}
    \hline
    Layers & Properties \\ \hline
       Dense  & size = 64, activation function = ReLU \\
       Dropout & dropout rate = $0.1$ \\
       Dense & size = 32, activation function = ReLU \\
       Dropout & dropout rate = $0.1$ \\
       Dense & activation function = ReLU \\
       Dropout & dropout rate = $0.1$ \\
       Dense & size = 2, activation function  = SoftMax \\ \hline
    \end{tabular}
    \end{subtable} 
    
\end{table}

\begin{table}[ht]
\small
    \centering
    \caption{Classification of Device 1 vs. Device 2 using legitimate LoRa device transmissions)}
    \label{tab:device12}
    \begin{subtable}[t]{0.5\textwidth}
    \centering
    \caption{DNN type: CNN.}
    \label{tab:actualCNN}
    \begin{tabular}{l||l}
    \hline
    Average accuracy of classifying  &0.9330  \\ Device 1 vs. Device 2 & \\ \hline
    Accuracy of detecting Device 1 & 0.9524 \\ (when Device 1 is transmitting) &  \\ \hline
    Accuracy of detecting Device 2 & 0.9133 \\ (when Device 2 is transmitting) &  \\ \hline 
    \end{tabular}
    \vspace{0.5cm}
    \end{subtable}
    \begin{subtable}[t]{0.5\textwidth}
    \centering
    \caption{DNN type: FNN.}
    \label{tab:actualFNN}
    \begin{tabular}{l||l}
    \hline
    Average accuracy of classifying  & 0.9180  \\ Device 1 vs. Device 2 & \\ \hline
    Accuracy of detecting Device 1 & 0.9345 \\ (when Device 1 is transmitting) &  \\ \hline
    Accuracy of detecting Device 2 & 0.9012  \\ (when Device 2 is transmitting) &  \\ \hline
    \end{tabular}
    \end{subtable} \label{tab:actual}
    
\end{table}

\section{Signal Spoofing by Rogue Devices} \label{sec:rogue}
For signal spoofing, the adversary uses KDE as a non-parametric technique to estimate the PDF of a random variable based on observed data. For each observed data point, a kernel function is centered at that point, and its contribution to the PDF estimate is calculated based on the chosen kernel and bandwidth. The kernel function determines the shape of the kernel. The bandwidth determines the width of the kernel, affects the smoothness, and balances between capturing fine details and avoiding oversmoothing of the estimated PDF. The kernel contribution is a scaled version of the kernel function evaluated at the given data point. Then, the individual kernel contributions are summed up to obtain the final estimated PDF. This summation process ensures that the estimated PDF is smooth and continuous. By applying KDE to observed data, the resulting estimate provides a representation of the underlying probability distribution. KDE allows for the generation of synthetic samples that follow the statistical properties of the observed data, making it a useful tool for generating synthetic signals that closely resemble legitimate LoRa signals. Given a set of observed data points $(x_1, x_2, \ldots, x_n)$, the KDE estimate $f(x)$ of the underlying PDF at a point $x$ is calculated as
\begin{equation}
f(x) = \frac{1}{nh} \sum_{i=1}^{n} K\left(\frac{x - x_i}{h}\right), \label{eq:KDE}
\end{equation}
where $K$ is the kernel function, $h$ is the bandwidth, $x_i$ represents each observed data point, and $n$ is the total number of observed data points. The KDE estimate $f(x)$ at a particular point $x$ is obtained by summing up the scaled contributions of the kernel function evaluated at each observed data point $x_i$. The scaling factor $\frac{1}{nh}$ ensures that the estimated PDF integrates to $1$ over the entire domain. 

\begin{figure}[t]
\captionsetup[subfigure]{justification=centering}
\centering
\begin{subfigure}[b]{0.49\columnwidth} 
\centering
\includegraphics[width=\columnwidth]{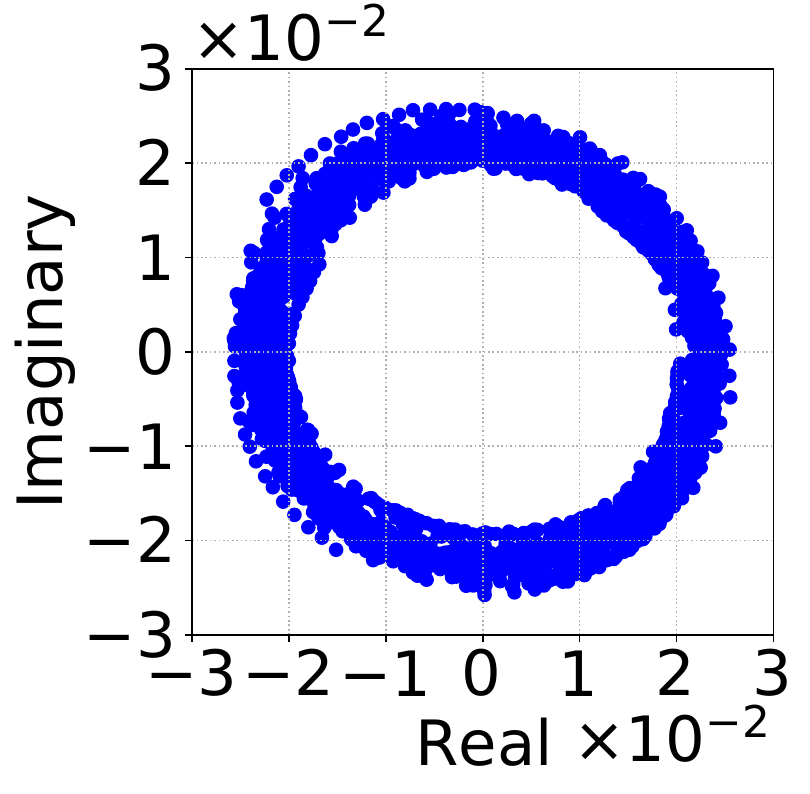}
\caption{Legitimate Device 1.}
\label{fig:Imagetrigger}
\end{subfigure}
\begin{subfigure}[b]{0.49\columnwidth}
\centering
\includegraphics[width=\columnwidth]{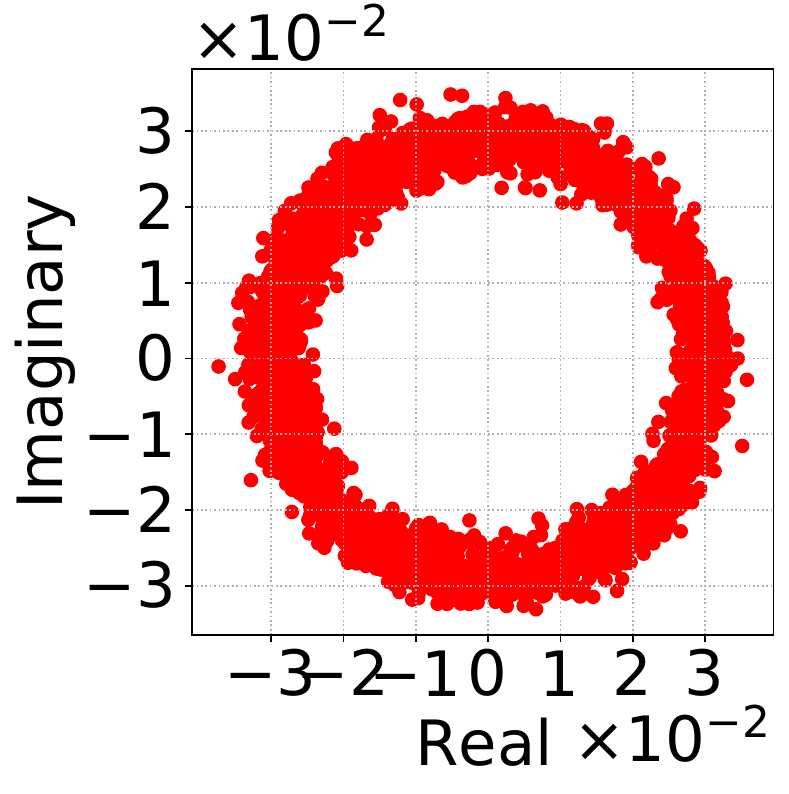}
\caption{Rogue Device 1.}
\label{fig:Imagerecons}
\end{subfigure}
\begin{subfigure}[b]{0.49\columnwidth}
\centering
\vspace{0.3cm}
\includegraphics[width=\columnwidth]{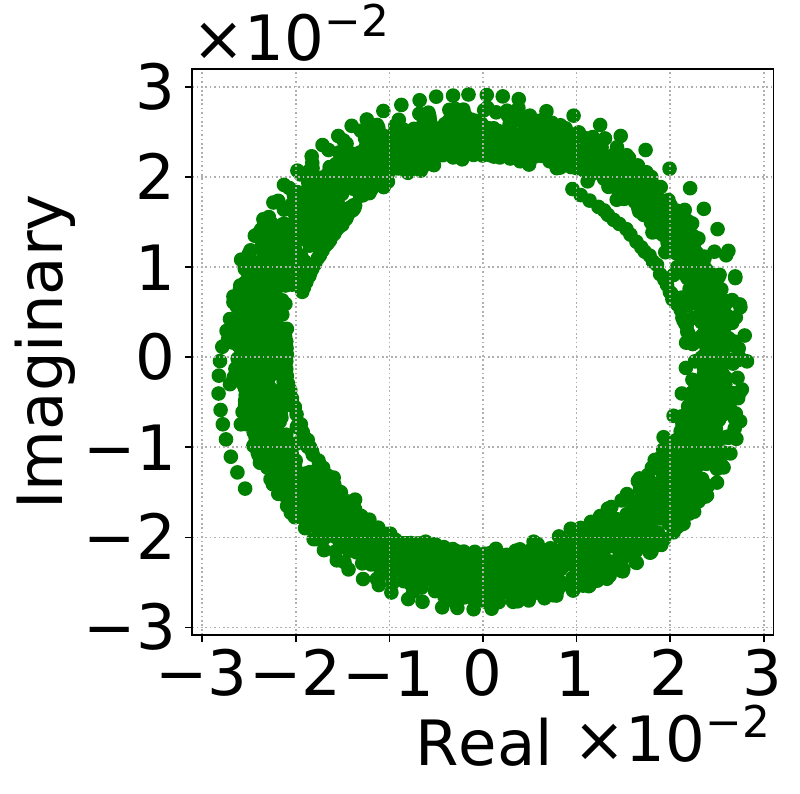}
\caption{Legitimate Device 2.}
\label{fig:Imagetrigger}
\end{subfigure}
\begin{subfigure}[b]{0.49\columnwidth}
\centering
\vspace{0.3cm}
\includegraphics[width=\columnwidth]{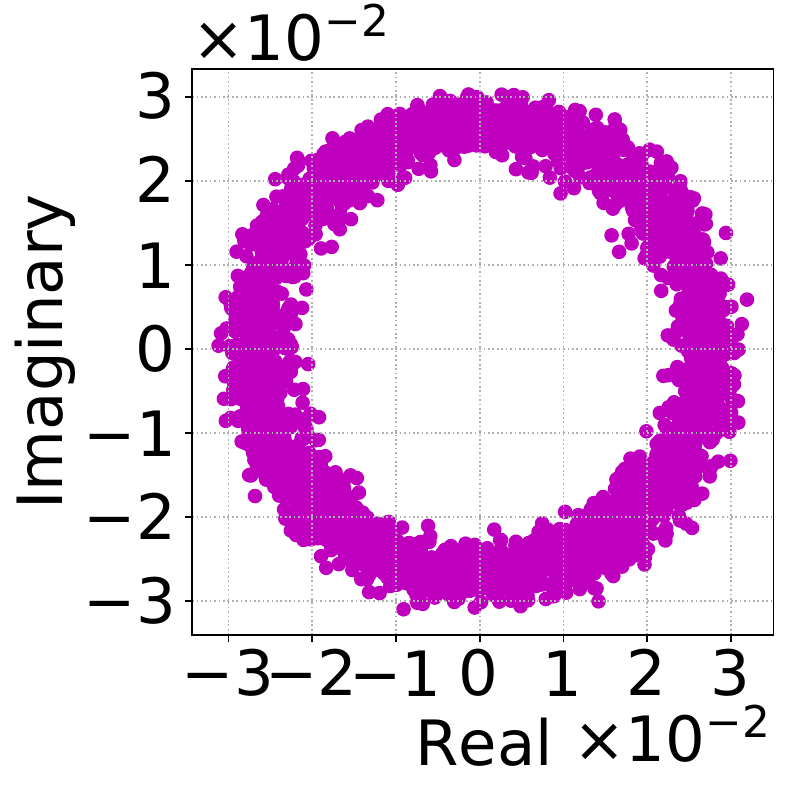}
\centering
\caption{Rogue Device 2.}
\label{fig:Imagerecons}
\end{subfigure}
\caption{Constellation of legitimate and rogue device signals (100 samples, where each sample consists of 32 I/Q signals).} \label{fig:constellation}
\end{figure}

To generate spoofed signals, we use Gaussian distribution as the kernel function and $10^{-3}$ as the bandwidth. Rogue Device 1 (that mimics legitimate Device 1) transmits with more than 2dB difference from legitimate Device 1, while rogue Device 2 (that mimics legitimate Device 2) transmits with less than 1dB difference from legitimate Device 2. Each rogue device has up to $\frac{\pi}{30}$ phase difference from the respective legitimate device. Fig.~\ref{fig:constellation} shows the constellation for the legitimate and rogue devices 1 and 2.  

We use the Jensen-Shannon divergence (JSD) to evaluate the fidelity of synthetic data generated by KDE. JSD quantifies the similarity between two probability distributions by calculating the average of the Kullback-Leibler (KL) divergences between each distribution and the average distribution. Given probability distributions $P_i$ and $\hat P_i$ for legitimate device and rogue device $i=1,2$, respectively, the JSD is computed as
\begin{equation}
\textit{JSD}(P_i, \hat P_i ) = \frac{1}{2}  \left(\textit{KL}(P_i || M_i ) +  \textit{KL}(\hat P_i || M_i ) \right),
\end{equation}
where 
\begin{equation}
\textit{KL}(P||Q) = \sum_{x \in \mathcal{X}} P(x) \log \left( \frac{P(x)}{Q(x)}\right)
\end{equation}
represents the KL divergence between discrete probability distributions $P$ and $Q$ defined on the same sample space $\mathcal{X}$, and $M_i = \frac{1}{2} \bigl(P_i+\hat P_i \bigl)$. The JSD ranges between 0 and 1, with 0 indicating that the two distributions are identical, and 1 indicating that the distributions are completely dissimilar. Therefore, a lower JSD value signifies a higher fidelity between the synthetic data generated by KDE and the original observed data. 
For test data, $\textit{JSD}(P_1, \hat P_1 ) = 0.0092$ for Device 1 and $\textit{JSD}(P_2, \hat P_2 ) = 0.0099$ such that the average JSD is $0.0096$ for both devices. This low JSD suggests that the spoofed signals closely resemble the original LoRa signals in terms of their statistical characteristics, indicating a high fidelity of LoRa signal spoofing by rogue devices. 

\begin{table}[]
\small
    \centering
    \caption{Classification accuracy using legitimate and rogue LoRa device transmissions when CNN is used as the DNN.}
    \label{tab:threstasksCNN}
    \begin{subtable}[t]{0.4\textwidth}
    \centering
    \caption{Task of classifying legitimate vs. rogue devices (using Device 1 and Device 2 transmissions).}
    \begin{tabular}{l||l}
    \hline
    Average accuracy of classifying  & 0.9755 \\ legitimate vs. rogue devices & \\ \hline 
    Accuracy of detecting the legitimate device & 0.9800 \\ 
    (when a legitimate device is transmitting) & \\ \hline
    Accuracy of detecting the rogue device & 0.9710 \\ 
    (when a rogue device is transmitting) & \\ \hline
    \end{tabular}
    \vspace{0.5cm}
    \end{subtable}
    
    \begin{subtable}[t]{0.4\textwidth}
    \centering
    \caption{Task of classifying Device 1 vs. Device 2 (using legitimate and rogue device transmissions).}
    \begin{tabular}{l||l}
    \hline
    Average Accuracy of classifying  & 0.9060 \\ Device 1 vs. Device 2 \\ \hline 
    Average accuracy of detecting Device 1 & 0.8946 \\ (when Device 1 is transmitting) & \\ \hline 
     Average accuracy of detecting Device 2 & 0.9175 \\ (when Device 2 is transmitting) & \\ \hline 
    \end{tabular}
    \vspace{0.5cm}
    \end{subtable} 

    \begin{subtable}[t]{0.4\textwidth}
    \centering
    \caption{Task of classifying Device 1 vs. Device 2 (using only rogue device transmissions).}
    \begin{tabular}{l||l}
     \hline
    Average Accuracy of classifying  & 0.9300 \\ Device 1 vs. Device 2 \\ \hline 
    Average accuracy of detecting Device 1 & 0.9226 \\ (when Device 1 is transmitting) & \\ \hline 
    Average accuracy of detecting Device 2 & 0.9375 \\ (when Device 2 is transmitting) & \\ \hline 
    \end{tabular}
    \end{subtable} 
    
\end{table}

\begin{table}[]
\small
    \centering
    \caption{Classification accuracy using legitimate and rogue LoRa device transmissions when FNN is used as the DNN.}
    \label{tab:threstasksFNN}
    \begin{subtable}[t]{0.4\textwidth}
    \centering
    \caption{Task of classifying legitimate vs. rogue devices (using Device 1 and Device 2 transmissions).}
    \begin{tabular}{l||l}
    \hline
    Average accuracy of classifying  & 0.9700 \\ legitimate vs. rogue devices & \\ \hline 
    Accuracy of detecting the legitimate device & 0.9771 \\ 
    (when an legitimate device is transmitting) & \\ \hline
    Accuracy of detecting the rogue device & 0.9629 \\ 
    (when a rogue device is transmitting) & \\ \hline
    \end{tabular}
    \vspace{0.5cm}
    \end{subtable}
    
    \begin{subtable}[t]{0.4\textwidth}
    \centering
    \caption{Task of classifying Device 1 vs. Device 2 (using legitimate and rogue device transmissions).}
    \begin{tabular}{l||l}
    \hline
    Average Accuracy of classifying  & 0.8956 \\ Device 1 vs. Device 2 \\ \hline 
    Average accuracy of detecting Device 1 & 0.8737 \\ (when Device 1 is transmitting) & \\ \hline 
     Average accuracy of detecting Device 2 & 0.9175 \\ (when Device 2 is transmitting) & \\ \hline 
    \end{tabular}
    \vspace{0.5cm}
    \end{subtable} 

    \begin{subtable}[t]{0.4\textwidth}
    \centering
    \caption{Task of classifying Device 1 vs. Device 2 (using only rogue device transmissions).}
    \begin{tabular}{l||l}
     \hline
    Average Accuracy of classifying  & 0.9190 \\ Device 1 vs. Device 2 \\ \hline 
    Average accuracy of detecting Device 1 & 0.9102 \\ (when Device 1 is transmitting) & \\ \hline 
    Average accuracy of detecting Device 2 & 0.9274 \\ (when Device 2 is transmitting) & \\ \hline 
    \end{tabular}
    \end{subtable} 
    
\end{table}

The DNN architectures shown in Table \ref{tab:NNArch} are also used for training classifiers to detect spoofed signals. In Section~\ref{sec:device}, we defined the task of classifying Device 1 vs. Device 2 using legitimate LoRa device transmissions. In this section, we consider three more tasks: 
task of classifying legitimate vs. rogue devices using Device 1 and Device 2 transmissions, task of classifying Device 1 vs. Device 2 from legitimate and rogue device transmissions, and task of classifying Device 1 vs. Device 2 using rogue device transmissions only. Tables \ref{tab:threstasksCNN} and \ref{tab:threstasksFNN} show the accuracy when CNN and FNN are used as the DNN, respectively. 
For each task, the accuracy is high both on average and for the case of detecting devices of individual labels (legitimate device vs. rogue device or Device 1 vs. Device 2). Overall, the accuracy is higher when devices are classified as legitimate device vs. rogue device compared to the case when devices are classified as Device 1 vs. Device 2. For the latter case of device classification, the accuracy drops when both legitimate and rogue device transmissions are used instead of using either legitimate or rogue device transmissions separately.

\section{Adversarial Attack} \label{sec:attack}
FGSM is an effective technique used for crafting adversarial examples in deep learning models. It is a one-step attack that generates adversarial perturbations by leveraging the gradient information of the loss function with respect to the input data. The FGSM attack aims to maximize the loss of the model by perturbing the input data in the direction of the gradient sign. 

The perturbation of the FGSM attack is generated by selecting an input sample to generate an adversarial example and calculating the gradient of the loss function with respect to the input data by backpropagating through the model. The gradient information determines the direction in which the input needs to be perturbed. This direction is computed as the sign of the gradient. Then, the gradient sign is scaled by a small magnitude ($\epsilon$) to control the strength of the perturbation. The value of $\epsilon$ determines the trade-off between the strength of the attack and the perceptibility of the perturbation. 

To launch the attack, the scaled perturbation is added to the original input sample. This is achieved by element-wise addition while ensuring that the perturbed input is within the permissible range of values for the data type controlled by the maximum transmit power of the adversary.
Given the DNN model with parameters $\theta$, an input sample $x$, and its true label $y$, the objective of the untargeted adversarial attack is to find a perturbation $\delta$ that maximizes the loss function while satisfying certain constraints. This optimization is written as
\begin{equation}
\max_{\delta} \mathcal{L}(x+\delta, y, \theta)  \label{eq:attack}
\end{equation}
subject to
\begin{enumerate}[label={C}{{\arabic*}}:]
\item $\|\delta\|_p \leq \epsilon_{max}$: the magnitude of perturbation $\delta$ is upper-bounded by $\epsilon_{max}$.
\item $x+\delta$ remains within the valid input range depending on the transmit power and phase shift of the device.
\end{enumerate}

As the non-convex optimization problem in (\ref{eq:attack}) is hard to solve, FGSM solves the optimization problem for an untargeted adversarial attack by linearizing the loss function with respect to the input perturbation. This linear approximation allows for an efficient computation of the perturbation that maximizes the loss. Specifically, the FGSM attack generates an adversarial example $x_{\text{adv}}$ by perturbing the input in the direction that maximizes the loss function with respect to the true label. The loss function $L(x, y, \theta)$ is computed for the input $x$ with respect to the true label $y$. The gradient of the loss function is computed with respect to the input as $\nabla_x L(x, y, \theta)$. This gradient is normalized by taking the sign of its elements as $\text{sign} (\nabla_x L(x, y, \theta))$. The sign of the gradient is scaled by a small value $\epsilon$, typically referred to as the step size or perturbation magnitude. Then, the perturbation for the untargeted attack is computed as 
\begin{equation}
\delta = \epsilon \: \text{sign}(\nabla_x L(x, y, \theta)),
\label{eq:singlepertubation}
\end{equation} 
where the $\text{sign}$ function extracts the sign of the gradient. The adversarial example $x_{\text{adv}}$ is generated by adding the perturbation to the original input such that 
\begin{equation}
x_{\text{adv}} = x + \epsilon \: \text{sign}(\nabla_x L(x, y, \theta)).    
\end{equation}
This perturbed signal \(x_{\text{adv}}\) is transmitted to fool the DNN model into making incorrect misclassification for any label. We define the classifier that classifies received signals as legitimate or rogue LoRa device as `Classifier 1', and the classifier that classifies received (legitimate or rogue) signals as LoRa Device 1 or Device 2 as `Classifier 2'. The attack success probability is the probability of misclassifying input signals without specifying any target label. We evaluate the attack success probability as a function of the perturbation-to-signal ratio (PSR) that determines the upper bound on the perturbation magnitude, $\epsilon$. 

\begin{figure}[t]
\vspace{-0.45cm}
	\centering
	\includegraphics[width=\columnwidth]{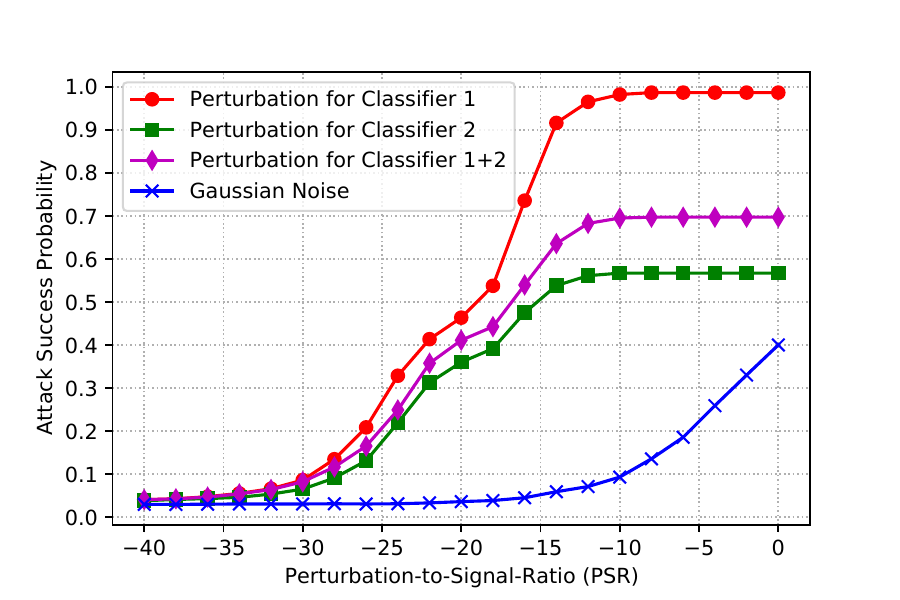}
	\caption{Untargeted attack on CNN classifier 1.}
	\label{fig:class1CNN}

	\centering
	\includegraphics[width=\columnwidth]{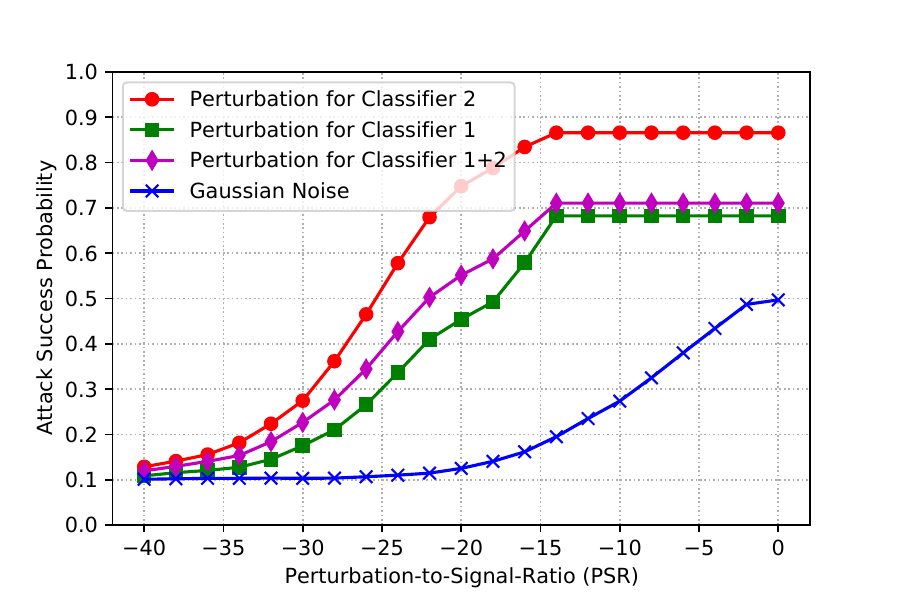}
	\caption{Untargeted attack on CNN classifier 2.}
	\label{fig:class2CNN}
 \vspace{-0.3cm}
\end{figure}

\begin{figure}[t]
\vspace{-0.45cm}
	\centering
	\includegraphics[width=\columnwidth]{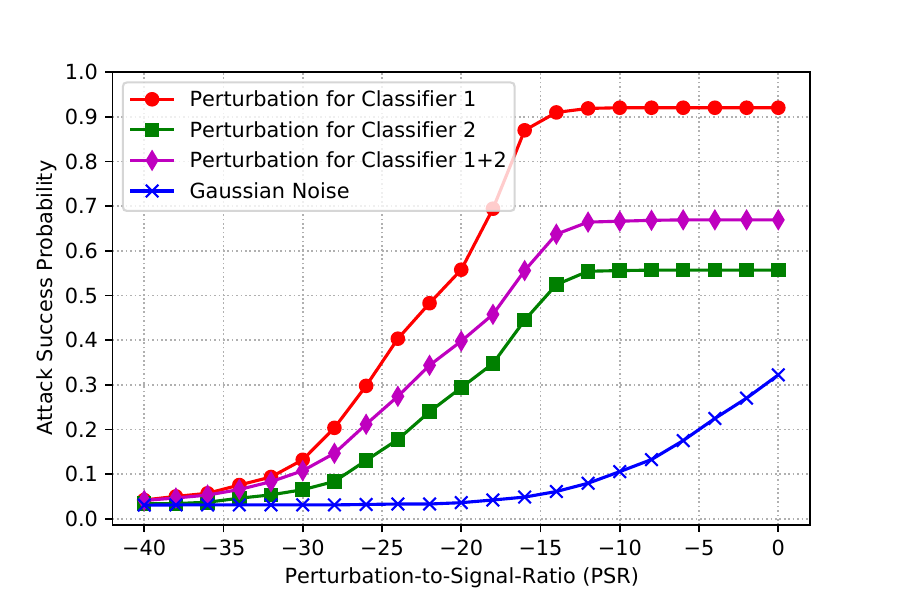}
	\caption{Untargeted attack on FNN classifier 1.}
	\label{fig:class1FNN}
 
	\includegraphics[width=\columnwidth]{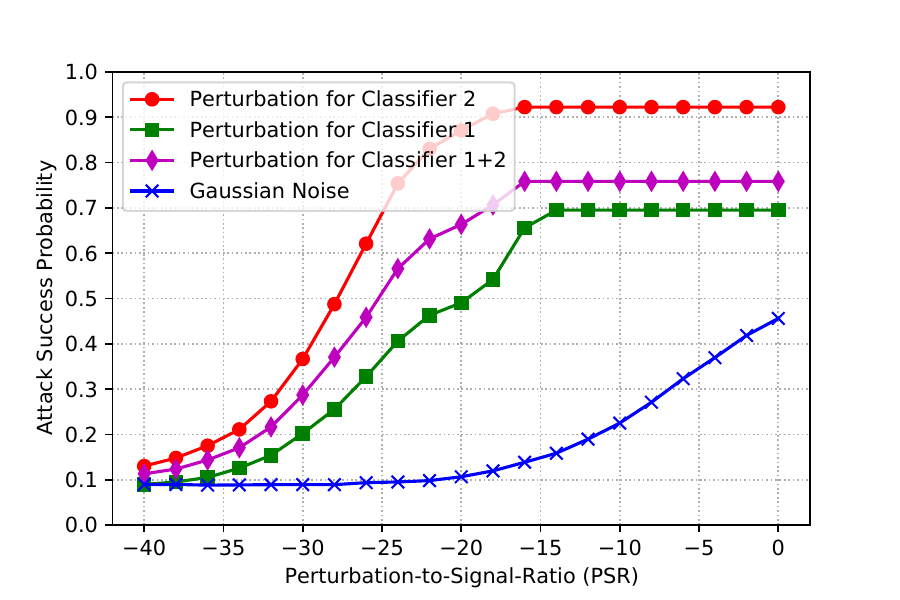}
	\caption{Untargeted attack on FNN classifier 2.}
	\label{fig:class2FNN}
  \vspace{-0.2cm}
\end{figure}

Fig.~\ref{fig:class1CNN} and Fig.~\ref{fig:class2CNN} show the attack success probability for the untargeted attack on CNN-based Classifier 1 and Classifier 2, respectively. Fig.~\ref{fig:class1FNN} and Fig.~\ref{fig:class2FNN} show the attack success probability for the untargeted attack on FNN-based classifier 1 and 2, respectively. We consider three cases of perturbation:
\begin{enumerate}
    \item Perturbation for Classifier 1: Perturbation is determined to maximize the loss of Classifier 1 according to (\ref{eq:singlepertubation}).
    \item Perturbation for Classifier 2: Perturbation is determined to maximize the loss of Classifier 2 according to (\ref{eq:singlepertubation}).
    \item Perturbation for Classifier 1+2: Perturbation is determined to maximize the weighted loss of Classifiers 1 and 2 according to
\begin{equation} \label{eq:hybrid}
\delta = \epsilon \: \text{sign} \biggl( \sum_{i=1}^2 w_i \nabla_x L_i(x, y_i, \theta_i) \biggl),
\end{equation} 
where $w_i$ is the weight for Classifier $i = 1,2$ (such that $0 \leq w_i \leq 1$ and $w_1+w_2= 1$) and $L_i$ is the loss function for classifier $i$ with parameters $\theta_i$. For numerical results, we set $w_1=w_2=0.5$. 
\end{enumerate}

We also evaluate the effectiveness of using Gaussian noise as perturbation signal as the benchmark. In this case, random perturbations are included to the signal independent of the input samples.

When launching an untargeted attack on a classifier using a perturbation designed specifically for that same classifier, the attack success probability is high for both Classifier 1 and 2. The best attack performance is achieved when the attack is launched on the CNN-based Classifier 1 and the perturbation is determined specifically for that classifier. However, if there is a mismatch between the attacked model and the model for which the perturbation was generated, the attack success probability diminishes. The extent of this decrease varies depending on the classifier under attack (with a substantial drop for Classifier 1 and a smaller drop for Classifier 2). With a hybrid attack (on `Classifier 1+2'), the effects of this mismatch are balanced in terms of attack transferability, enabling the adversary to employ a common perturbation that is broadcast (with a single transmission) to achieve high attack success against each classifier. In all cases, the use of Gaussian noise as perturbation signal for brute-force jamming is ineffective in reducing the classifier accuracy significantly. When we compare the performance of CNN classifier vs. FNN classifier, we observe that attack success probability is slightly higher for attacks designed against the CNN classifier.

\section{Conclusion} \label{sec:conclusion}
We addressed the security concerns associated with LoRa networks that provide long-range and low-power communication capabilities for IoT applications. We developed a deep learning framework for two tasks of device identification and classification of legitimate and spoofed signals in LoRa networks. By employing CNN or FNN as the DNN for these tasks and using real experimental I/Q data for LoRa signals, along with KDE for generating spoofed signals by rogue devices, we studied the effectiveness of FGSM-based untargeted adversarial attacks on these LoRa signal classification tasks. We showed that these attacks are highly effective in reducing the classifier accuracy especially when the perturbation is determined for the particular classifier under attack. In addition,  we found out that a common perturbation can be effectively crafted by the adversary to achieve high attack success against each classifier simultaneously. Our results provided insights into the transferability of adversarial attacks, emphasized the vulnerability of LoRa networks to such attacks in IoT systems, and highlighted the need of defending against these attacks.

\bibliographystyle{IEEEtran}
\bibliography{references}

\end{document}